\documentclass{article}

\usepackage[dblblindworkshop, final]{neurips_2026}
\workshoptitle{E-Values: From Statistics to ML}

\usepackage[utf8]{inputenc}
\usepackage[T1]{fontenc}
\usepackage{float}
\usepackage{hyperref}
\usepackage{url}
\usepackage{booktabs}
\usepackage{multirow}
\usepackage{amsfonts}
\usepackage{amsmath}
\usepackage{amssymb}
\usepackage{nicefrac}
\usepackage{placeins}
\usepackage{xcolor}
\usepackage{graphicx}

\DeclareMathOperator*{\argmax}{arg\,max}
\newcommand{\GLS}{\mathrm{GLS}}

\newcommand{\Ind}{\mathbb{1}}

\newcommand{\E}{\mathbb{E}}
\newcommand{\Prob}{\mathbb{P}}

\newcommand{\deltafa}{\delta_{\mathrm{FA}}}

\title{Anytime-valid detection of LLM weight exfiltration}

\author{%
  Ines Ortega-Fernandez \\
  MATS Research \\
  \texttt{ines.ortega@uvigo.gal} \\
  \And
  Mateusz Kowalczyk \\
  NASK -- National Research Institute, MATS Research \\
  \texttt{mateusz.kowalczyk@nask.pl} \\
  \And
  Keri Warr \\
  Anthropic \\
  \texttt{keri@anthropic.com} \\
}

\usepackage{caption}
\usepackage{titlesec}
\titlespacing*{\section}{0pt}{8pt plus 2pt minus 2pt}{4pt}
\titlespacing*{\subsection}{0pt}{6pt plus 2pt minus 2pt}{3pt}
\renewcommand{\arraystretch}{0.95}
\AtBeginDocument{%
  \setlength{\abovedisplayskip}{4pt plus 2pt minus 2pt}%
  \setlength{\belowdisplayskip}{4pt plus 2pt minus 2pt}%
  \setlength{\abovedisplayshortskip}{2pt plus 1pt minus 1pt}%
  \setlength{\belowdisplayshortskip}{2pt plus 1pt minus 1pt}%
}

\begin{document}

\maketitle
\raggedbottom

\begin{abstract}
A compromised LLM inference server can leak model weights by encoding payload bits in otherwise plausible token choices. A replay of the same prompt in a trusted server can expose such deviations, but benign numerical nondeterminism also causes token mismatches. Patient attackers can therefore hide within normal variation unless evidence is combined across responses. We introduce a prompt-level e-process that calibrates whole-response mismatch events on trusted benign traffic and accumulates evidence sequentially while, under a calibration-transfer assumption, controlling the probability of any false alarm over an unbounded monitoring horizon. We evaluate it on four models against a seed-blind attack and a stronger seed-aware attack that hides payload bits only in near-ties to remain stealthy, analyzing the channel capacity vs detectability trade-off. Compared with a hard per-token alarm, the e-process combines weak evidence across responses while providing explicit anytime false-alarm control.
\end{abstract}

\section{Introduction}
\label{sec:intro}

An LLM inference server holds the model weights and runs the sampler that turns logits into tokens. If compromised, it can leak model weights, private prompts, or arbitrary data by using different plausible next tokens to represent payload bits in ordinary-looking responses~\citep{meteor2021,imec2023}. An inference server must return tokens to its clients, so the output-token channel can be monitored and rate-limited, but not completely closed~\citep{rinberg2025}. However, the model weights owner has an advantage: it can replay the same prompt in a trusted server (the \emph{verifier}) and check whether the served token matches the one prescribed by the committed sampling seed. The sampling seed determines one prescribed token $Y^*$ at every position, so a different served token is a \emph{token mismatch}. Under Gumbel--Max sampling~\citep{gumbel1954statistical}, the fixed seed determines one Gumbel offset per logit, and the largest perturbed logit defines $Y^*$ (Eq.~\ref{eq:gumbel}). These offsets define the prescribed sampling outcome, but small numerical differences due to GPU non-associativity can reorder a near-tie under the same offsets and cause a benign token mismatch~\citep{karvonen2025}. Seed access therefore determines whether an attacker can exploit this cover to remain stealthy. A \emph{seed-blind} attacker sees the model logits but not the offsets, so it cannot know $Y^*$ or the gap to the winner (the seeded margin). It must choose a token to carry the payload before knowing how the offsets reorder the candidates, so its mismatches need not fall in the near-ties where benign token mismatches concentrate. A \emph{seed-aware} attacker can reconstruct both $Y^*$ and the seeded margin. It can restrict payload-carrying tokens to seeded near-ties and return $Y^*$ elsewhere, making deliberate mismatches look like benign ones. Appendix~\ref{subsec:threat-model} gives more details about the attack implementation, the threat model, and their decoding assumptions.

Existing detection rules classify individual tokens from their fixed-seed score and raw-logit rank~\citep{rinberg2025}, but leave response-level pooling and escalation to deployment. A patient attacker may leave too little evidence in any one token, yet repeatedly applying a hard token alarm eventually flags benign traffic unless error is controlled over the whole monitoring path. We close this gap with an e-process~\citep{ramdas2023} over whole-response (rather than token-level) mismatch events: under the prompt-level null and a calibration-transfer assumption, it runs indefinitely while keeping the probability of any false alarm below a predefined rate. On four models, it detects every resampled seed-blind stream and several seed-aware settings. By contrast, the hard-alarm instantiation of \citet{rinberg2025} can't detect the seed-blind stream while alarming on nearly every benign stream.

\textbf{Related work}. Our monitor builds on two closely related works. \citet{rinberg2025} formalize weight exfiltration through the sampler and bound its capacity under a fixed-seed verifier, with a per-token classifier (safe, suspicious, or dangerous) read from a per-token fixed-seed likelihood score and a raw-logit rank. Token-DiFR~\citep{karvonen2025} takes up the complementary difficulty that replay is not bit-exact, characterizing when a mismatch between served and replayed tokens is still consistent with benign serving under numerical nondeterminism. Our monitor takes the fixed-seed score from the first and the benign-mismatch analysis from the second. Methodologically, \citet{shin2024} construct e-detectors from restarted e-processes for general sequential change detection, while \citet{su2026} use e-processes for online LLM watermark detection. Neither addresses inference-server weight exfiltration. A companion paper~\citep{ortega2026conformal} studies the complementary calibration question on the same threat model and corpus: it equalizes the realized per-token false-alarm rate across entropy strata and adds a fixed-horizon per-response test, which controls error for a single look rather than over the unbounded horizon considered here.

\section{Method: a prompt-level e-process}
\label{sec:method}

The construction starts from $K$ nested binary events $A_{j,m}$ computed after response $j$. Each event must be fixed before its benign rate is calibrated. For token $t$ in response $j$, the verifier recomputes the token prescribed by the committed seed, using Gumbel--Max sampling~\citep{gumbel1954statistical}, and records whether the server returned a different token:
\begin{equation}
Y^*_{j,t}=\argmax_i\{l_{j,t,i}/T+g_{j,t,i}\},
\qquad F_{j,t}=\Ind\{y_{j,t}\ne Y^*_{j,t}\}.
\label{eq:gumbel}
\end{equation}
A benign token mismatch is most plausible when the two leading perturbed logits are nearly tied. Let $\Delta_{j,t}$ be the gap between these logits. An independent benign split fixes ordered thresholds $0=b_0<\cdots<b_{K-1}$. The margin event is
\begin{equation}
A_{j,m}=\Ind\{\exists t:\ F_{j,t}=1\ \text{and}\ \Delta_{j,t}\ge b_m\},
\qquad m=0,\ldots,K-1.
\label{eq:prompt-event}
\end{equation}
A higher-threshold event also triggers every lower-threshold event. Each response contributes one binary observation per threshold, so no independence is assumed among its tokens.

\paragraph{Benign calibration.} After fixing the events, we use $n$ new benign calibration responses. Let $A^{\mathrm{cal}}_{i,m}$ be event $m$ on response $i$, and let $k_m=\sum_{i=1}^{n}A^{\mathrm{cal}}_{i,m}$. Thus, $k_m$ counts calibration responses that trigger event $m$, not individual tokens. The plain fraction $k_m/n$ can underestimate the rate by chance, so we instead use the Clopper--Pearson upper bound $q_m^+$~\citep{clopper1934}, fixing $q_m^+$ such that the probability of observing at most $k_m$ events under a possible rate $q_m^+$ reaches $\eta=\gamma/K$:
\begin{equation}
\sum_{r=0}^{k_m}\binom{n}{r}(q_m^+)^r(1-q_m^+)^{n-r}=\eta,
\qquad k_m<n.
\label{eq:calibration}
\end{equation}
We set $q_m^+=1$ when $k_m=n$. Equation~\eqref{eq:calibration} asks how large the benign rate could be while observing only $k_m$ or fewer events remains an $\eta$-probability outcome. Each bound misses the true rate with probability at most $\eta$, so all $K$ bounds cover their rates together with probability at least $1-\gamma$. A loose bound may delay detection, but cannot make benign responses appear more suspicious.

The monitor then asks whether future responses continue to trigger each event at no more than the benign rate $q_m^+$. Let $\mathcal G_{j-1}$ contain calibration and the completed prompt-response rounds through $j-1$, before the next prompt-response round is drawn. We formalize the null hypothesis as
\begin{equation}
\begin{aligned}
H_0:&\quad \E_0[A_{j,m}\mid\mathcal G_{j-1}]\le q_m
&&\text{for every response $j$ and threshold index $m$}.
\end{aligned}
\label{eq:null}
\end{equation}
Under $H_0$, conditional on past rounds, the next benign response triggers each event no more often than its benign rate. This guarantee averages over prompts from the calibration distribution, so it does not need to hold for each prompt separately. Under the alternative $H_1$, at least one event becomes more likely. Thus, $H_0$ requires the same prompt distribution, model, and serving stack used during trusted calibration. Prompt shift, adaptive prompt selection, or a serving-stack change may require stratified calibration or recalibration.

\paragraph{Sequential evidence.} For each threshold index $m$, the monitor starts with evidence equal to one and updates it after every response by multiplying by a factor:
\begin{equation}
E_{j,m}=1+\lambda_{j,m}(A_{j,m}-q_m^+),\quad
M_{J,m}=\prod_{j=1}^{J}E_{j,m}.
\label{eq:eprocess-component}
\end{equation}
An event raises the factor, and a response without it lowers it, so a sustained excess across responses accumulates evidence. We choose $\lambda_{j,m}$ from earlier responses only, and set it to zero unless their estimated event rate exceeds $q_m^+$, and cap it at $c/q_m^+$. With $c=0.5$, no update can multiply the evidence by less than $0.5$. Appendix~\ref{app:validity} gives the exact rule and its motivation. Finally, we combine the $K$ components using the fixed average
\begin{equation}
M_J=\frac{1}{K}\sum_{m=0}^{K-1}M_{J,m}.
\label{eq:eprocess}
\end{equation}

Validity follows from two facts. First, $\lambda_{j,m}$ is fixed before response $j$ is observed, so it cannot depend on that response's outcome. Second, successful calibration gives $\E_0[A_{j,m}\mid\mathcal G_{j-1}]\le q_m^+$. Together these imply $\E_0[E_{j,m}\mid\mathcal G_{j-1}]\le1$. The cap gives $E_{j,m}\ge1-c>0$. Conditional on successful calibration, each component is therefore a nonnegative supermartingale, and their fixed average is an e-process~\citep{ramdas2023}. We alarm when $M_J\ge1/\alpha$. Ville's inequality~\citep{ville1939} controls the probability of ever crossing this threshold, regardless of when monitoring stops. The conditional crossing probability is at most $\alpha$. Including calibration failure $\gamma$ gives
\begin{equation}
\Prob_0\!\left(\sup_{J\ge1}M_J\ge1/\alpha\right)
\le\alpha+\gamma=:\deltafa.
\label{eq:ville}
\end{equation}
Appendix~\ref{app:validity} provides the full proof.

\section{Experiments}
\label{sec:experiments}

We study two attack regimes distinguished by seed access (Appendix~\ref{subsec:threat-model}), with codecs tailored to the information available. The lower-knowledge \emph{seed-blind} attacker cannot reconstruct the prescribed token, so it uses arithmetic coding over clean near-ties and its mismatches land wherever the seeded perturbation falls. The stronger \emph{seed-aware} attacker reconstructs the seed and uses a binary choice between the prescribed token and its runner-up. Its parameter $d_{\max}$ caps the fixed-seed score gap it will exploit: increasing $d_{\max}$ makes more positions eligible for encoding bits but exposes mismatches at wider, more detectable seeded margins. 

For each model, $125$ benign responses fit seven margin events and another $125$ calibrate their rates. We use $c=0.5$ for the update cap and $\alpha=\gamma=0.005$, giving the operational e-process an alarm threshold of $200$ and a $1\%$ overall false-alarm bound. We compare it with the three-tier classifier of \citet{rinberg2025}, using their fixed-seed likelihood threshold $\tau=-5$ and raw-logit rank threshold $R=8$ (see Appendix~\ref{subsec:rinberg}). Their policy does not specify pooling over time, so we alarm at the first response containing a dangerous token. Appendix~\ref{subsec:dataset} specifies data splits, calibration, and detector settings.

Table~\ref{tab:compromise} reports how quickly each method alarms over $1{,}000$ bootstrap streams of $250$ responses. For each stream we record the response at first alarm and whether it alarms by response $250$.

\begin{figure}[ht]
  \centering
  \includegraphics[width=0.92\linewidth]{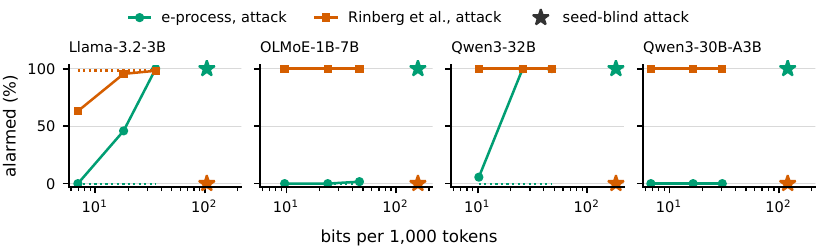}
  \caption{Fraction of streams alarmed by response $250$ ($1{,}000$ bootstrap streams).}
  \label{fig:capacity-detection}
\end{figure}

\begin{table}[t]
  \centering
  \caption{Time to alarm for seed-aware and seed-blind attacks for each detector, over $1{,}000$ bootstrap streams of $250$ responses. ``resp.\ to alarm'': median [$10$--$90$th pct] response over streams that alarm ($-$ if none by $250$).}
  \label{tab:compromise}
  \scriptsize
  \setlength{\tabcolsep}{5pt}
  \renewcommand{\arraystretch}{0.88}
  \begin{tabular}{@{}llrcrc@{}}
    \toprule
    & & & \multicolumn{2}{c}{e-process} & Rinberg et al. \\
    \cmidrule(lr){4-5}\cmidrule(lr){6-6}
    Model & $d_{\max}$ & bits/tok & resp.\ to alarm & \% alarm (attack\,/\,benign) & \% alarm (attack\,/\,benign) \\
    \midrule
    \multicolumn{6}{@{}l}{\emph{Seed-blind} attack (no seed access)} \\
    \midrule
    Llama-3.2-3B &  & .1045 & 6 [5--7] & 100.0\,/\,0.0 & 0.0\,/\,100.0 \\
    OLMoE-1B-7B &  & .1587 & 10 [9--13] & 100.0\,/\,0.0 & 0.0\,/\,100.0 \\
    Qwen3-32B &  & .1831 & 9 [9--9] & 100.0\,/\,0.0 & 0.0\,/\,100.0 \\
    Qwen3-30B-A3B &  & .1203 & 14 [10--21] & 100.0\,/\,0.0 & 0.0\,/\,100.0 \\
    \midrule
    \multicolumn{6}{@{}l}{\emph{Seed-aware} attack by $d_{\max}$} \\
    \midrule
    \multirow{3}{*}{Llama-3.2-3B} & .02 & .0069 & $-$ & 0.0\,/\,0.0 & 63.1\,/\,98.4 \\
        & .05 & .0182 & 172 [72--235] & 45.9\,/\,0.0 & 95.5\,/\,98.4 \\
        & .1 & .0357 & 28 [14--49] & 100.0\,/\,0.0 & 98.2\,/\,98.4 \\
    \midrule
    \multirow{3}{*}{OLMoE-1B-7B} & .02 & .0096 & $-$ & 0.0\,/\,0.0 & 100.0\,/\,100.0 \\
        & .05 & .0238 & $-$ & 0.0\,/\,0.0 & 100.0\,/\,100.0 \\
        & .1 & .0464 & 181 [115--249] & 1.7\,/\,0.0 & 100.0\,/\,100.0 \\
    \midrule
    \multirow{3}{*}{Qwen3-32B} & .02 & .0103 & 204 [152--243] & 5.6\,/\,0.0 & 100.0\,/\,100.0 \\
        & .05 & .0258 & 85 [67--116] & 100.0\,/\,0.0 & 100.0\,/\,100.0 \\
        & .1 & .0474 & 30 [26--39] & 100.0\,/\,0.0 & 100.0\,/\,100.0 \\
    \midrule
    \multirow{3}{*}{Qwen3-30B-A3B} & .02 & .0068 & $-$ & 0.0\,/\,0.0 & 100.0\,/\,100.0 \\
        & .05 & .0165 & $-$ & 0.0\,/\,0.0 & 100.0\,/\,100.0 \\
        & .1 & .0304 & $-$ & 0.0\,/\,0.0 & 100.0\,/\,100.0 \\
    \bottomrule
  \end{tabular}
\end{table}

We begin with the seed-blind attacker (Table~\ref{tab:compromise}, top block). The e-process alarms on all $1{,}000$ resampled attack streams for every model at a median of $6$--$14$ responses, while no matched benign stream ever alarms. On the other hand, the method from \citet{rinberg2025} alarms only on benign streams because of its rank gate: a token is dangerous only if it is both unlikely under the committed seed and has raw-logit rank greater than $8$. The seed-blind encoder emits the clean maximum or a near-tie, so its many token mismatches remain within the top-rank and are labeled suspicious, not dangerous. However, under benign non-determinism, Gumbel sampling can instead prescribe a lower-ranked token, and a replay disagreement there satisfies both gates, flagging tokes as dangerous on the benign held-out set, alarming on every benign stream while being unable to detect the attack.

Seed access improves the attacker's position by letting it reconstruct the prescribed token and confine deviations to seeded near-ties, trading exfiltration capacity for stealth. Again, the e-process never false-alarms on benign traffic. Regarding attack detection, on the dense models the attack is detected at the higher evaluated rates. On the two mixture-of-experts models, benign mismatches occur more often at wide seeded margins, raising the calibrated event bounds and leaving little excess for near-tie attack mismatches to contribute (Appendix~\ref{app:margin}). This is an empirical property of these model--serving stacks, not evidence that mixture-of-experts architecture directly influences detectability. Detection therefore tracks both channel intensity and the benign margin profile (Figure~\ref{fig:capacity-detection}): Qwen3-32B is caught at $0.0258$ bits/token, while Qwen3-30B-A3B is not at the higher rate $0.0304$. More detailed experimental results and the evidence paths are available at Appendix~\ref{app:detailed-results}.

\section{Conclusion}
Inference verification can limit the exfiltration channel but cannot remove it, since any inference server has to produce responses. This work contributes a prompt-level e-process whose guarantee makes inference verification usable in practice: it can be checked after every response while keeping the probability of any false alarm below a preset level over an unbounded horizon. Across four models it alarms rapidly on every resampled seed-blind stream, while seed-aware power varies with rate and the benign margin profile of the model. Some seed-aware attacks that transmit only a few hundredths of a bit per token are not detected within $250$ responses. Such rates could slow enough bulk weight exfiltration but do not prevent the exfiltration of small secrets. Future work will explore an e-process built directly on the fixed-seed likelihood score instead of the binary margin events, betting on this continuous score to accumulate evidence faster and detect a seed-aware attacker at even lower rates.

\begin{ack} 
This work was conducted as part of the ML Alignment \& Theory Scholars (MATS) Program.
\end{ack}

{\small
\bibliographystyle{plainnat}
\bibliography{references}
}

\appendix

\section{Why the response statistic matters}
\label{app:margin}

The e-process can control false alarms for any prespecified response events whose calibrated bounds transfer to monitoring. The simplest event, ``this response contains at least one token mismatch'', is nearly useless because benign serving already produces a mismatch in $41.2\%$--$87.2\%$ of complete responses (Table~\ref{tab:benign-flips}). The e-process still controls its false-alarm probability, but it has little room to detect an attacker on top of that. An event that fires this often on benign traffic has no power against an attacker that exactly matches its benign event rate.

\begin{figure}[H]
  \centering
  \includegraphics[width=\linewidth]{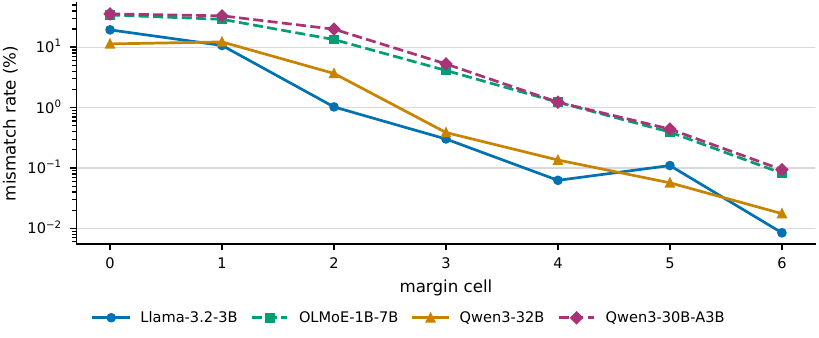}
  \caption{Per-token benign mismatch rate by margin cell in the same $250$ held-out benign responses as Table~\ref{tab:benign-flips}. Margin-cell edges were fitted on a separate $125$-response benign split. Cell $0$: tightest ties, cell $6$: widest margins.}
  \label{fig:margin-law}
\end{figure}

The solution is to track not only whether a mismatch occurred, but also \emph{where} it occurred. The fixed seeded Gumbel draws define the prescribed token and its margin. Benign token mismatches arise when numerical differences between the serving and verification stacks reorder a near-tie under those same draws. A mismatch at a wide seeded margin is therefore harder to explain by numerical noise alone. Figure~\ref{fig:margin-law} shows how steeply the benign mismatch rate falls as the margin widens. Each token falls in exactly one margin cell, and the curve is the fraction of tokens in that cell that mismatch: across models it drops from roughly $10\%$--$40\%$ in the tightest cell to below $0.1\%$ in the widest. These per-token rates are descriptive only: the monitor holds its guarantee by calibrating whole-response events on separate benign responses.

\begin{table}[H]
  \centering
  \caption{Benign nondeterminism in $250$ held-out benign responses across model types. The first column counts the fraction of tokens that mismatch the seeded replay, the second counts the fraction of complete responses that contain at least one mismatch, and the third counts the mean number of mismatches per response.}
  \label{tab:benign-flips}
  \footnotesize
  \setlength{\tabcolsep}{5pt}
  \begin{tabular}{@{}llrrr@{}}
    \toprule
    Model & Type & Token mismatches (\%) & Responses with mismatch (\%) & Mean mismatches/response \\
    \midrule
    Llama-3.2-3B  & Dense & 0.40 & 41.2 & 0.83 \\
    OLMoE-1B-7B   & MoE & 1.75 & 87.2 & 5.54 \\
    Qwen3-32B     & Dense & 0.55 & 79.2 & 2.46 \\
    Qwen3-30B-A3B & MoE & 1.65 & 74.8 & 4.61 \\
    \bottomrule
  \end{tabular}
\end{table}

The ordering does not track model scale, and we did not evaluate task capability against replay stability. OLMoE-1B-7B has the highest pooled mismatch burden ($1.75\%$ of tokens, $87.2\%$ of responses, and $5.54$ mismatches per response), while the larger dense Qwen3-32B is lower on all three measures. More broadly, both mixture-of-experts models have roughly $3$--$4$ times the dense models' pooled token mismatch rates and higher mismatch rates in the widest cell. Sparse expert routing is a plausible amplifier: small numerical changes may alter the selected experts and propagate into larger logit differences. Four checkpoints confound model family, architecture, and serving kernels, however, so this is a hypothesis rather than a capability trend or causal architecture result. We therefore treat benign noise as a property of the complete model--serving stack and calibrate each deployment separately.

A higher benign mismatch rate in a wide margin cell gives the attacker cover: payload-carrying mismatches look more like benign ones, and each adds less evidence. In the extreme, a monitor that ignores the margin entirely sees only mismatch counts, so an attacker that matches the overall benign rate is indistinguishable from benign traffic to that statistic. Conditioning on the margin adds information the seed-blind attack cannot target. A seed-aware attacker can better match where benign mismatches fall, but only by restricting the evaluated encoder to tighter ties and therefore fewer eligible positions.

In addition, a single margin threshold would force us to decide in advance whether to watch for common tight-margin excesses or for rare wide-margin ones. Instead, we run one e-process per threshold and average the $K=7$ of them, letting the stream itself determine where evidence accumulates. The price of not choosing is small: the average is never less than $1/K$ times the best component, a penalty of at most $\log K$ in log evidence, and no threshold is ever selected using attack data. The events are deliberately simple: each event records only whether it fired, not how many mismatches qualified or how wide their margins were. A richer response statistic could detect more, but it would need its own calibration and its own validity argument, that we leave for future work.

\section{Validity of the e-process}
\label{app:validity}

Let $\mathcal C_{\mathrm{edge}}$ and $\mathcal C_{\mathrm{event}}$ be independent sets of benign responses. The first fixes the margin thresholds and therefore the event maps in \eqref{eq:prompt-event}. The second estimates their probabilities. Let $\mathcal G_{j-1}$ contain both calibration sets and the completed monitoring rounds through $j-1$, before the next prompt-response round is drawn. Write $\mathcal C=(\mathcal C_{\mathrm{edge}},\mathcal C_{\mathrm{event}})$ for the combined calibration realization. The monitor updates only after a response is complete. There are $K$ nested events, indexed by $m=0,\ldots,K-1$.

\paragraph{Assumptions.}
\begin{enumerate}
  \item Conditional on $\mathcal C_{\mathrm{edge}}$, the responses in $\mathcal C_{\mathrm{event}}$ are independent draws from the benign response distribution. For each $m$, their indicators $A^{\mathrm{cal}}_{i,m}$ are therefore Bernoulli with mean $q_m$.
  \item During monitoring, the prompt-level null \eqref{eq:null} holds for every round and threshold index. Independent prompt-response rounds from the calibrated deployment distribution are sufficient, but not necessary.
  \item Calibration and monitoring use the same prompt distribution, model, hardware, and serving stack, and the calibration responses are collected from a trusted window in which the server is known benign. Because the threat is a compromised serving stack, this trusted window is what the guarantee is measured against: it bounds false alarms on deviations that occur after calibration. The expectation in \eqref{eq:null} is marginal over the next prompt and response. The calibration does not establish a uniform bound conditional on every possible prompt: prompt shift or adaptive prompt selection can violate the null.
  \item The coefficient $\lambda_{j,m}$ is $\mathcal G_{j-1}$-measurable and satisfies $0\le\lambda_{j,m}\le c/q_m^+$ for a fixed $c<1$.
\end{enumerate}

\paragraph{Choosing $\lambda$ from past responses.} Let $S_{j-1,m}=\sum_{i<j}A_{i,m}$ count the earlier responses that triggered event $m$. Before observing response $j$, we add half a count ($1/2$) to each outcome: event and no event. The resulting estimate is
\[
\hat r_{j,m}=\frac{S_{j-1,m}+1/2}{(S_{j-1,m}+1/2)+(j-1-S_{j-1,m}+1/2)}=\frac{S_{j-1,m}+1/2}{j}.
\]
This is the Krichevsky--Trofimov estimator~\citep{krichevsky1981}: the two half counts act as a weak prior observation of each outcome, so no finite sample can drive the estimate to exactly zero or one. Without the half counts, the first few responses alone would push the estimate to an extreme: zero after a quiet start with no events, or one after an early event, and the coefficient below would swing with it. Applying the smoothing allows early estimates to stay moderate, and the coefficient can adjust gradually as responses accumulate. The coefficient turns this estimate into the size of the next update: it grows with the excess of the estimated rate over the calibrated bound, and it is clipped to the permitted range $[0,c/q_m^+]$:
\begin{equation}
\widetilde{\lambda}_{j,m}=\frac{\hat r_{j,m}-q_m^+}{q_m^+(1-q_m^+)},\qquad \lambda_{j,m}=\min\!\left\{\frac{c}{q_m^+},\max\!\left\{0,\widetilde{\lambda}_{j,m}\right\}\right\}.
\label{eq:lambda}
\end{equation}
If $q_m^+=1$, we set $\lambda_{j,m}=0$. Because this choice uses only earlier responses, $\lambda_{j,m}$ is fixed before response $j$ is observed. This is the predictability condition required for validity. To see where the formula comes from, write $q$ for $q_m^+$ and suppose the next event occurs with probability $r$. The expected logarithm of the update factor is
\begin{equation}
g(\lambda;r,q)=r\log\!\left(1+\lambda(1-q)\right)+(1-r)\log\!\left(1-\lambda q\right).
\label{eq:expected-log-growth}
\end{equation}
Differentiating \eqref{eq:expected-log-growth} gives the growth-rate-optimal fixed bet $\lambda^*=(r-q)/[q(1-q)]$~\citep{wsr2023}. Since $r$ is unknown, substituting the past-response estimate $\hat r_{j,m}$ gives $\widetilde{\lambda}_{j,m}$ in \eqref{eq:lambda}. Clipping at zero prevents the monitor from reacting when the estimated rate is below the benign bound. The upper limit $c/q_m^+$ ensures that every update factor is at least $1-c$. The estimate need not be correct for validity, $\lambda_{j,m}$ only needs to be chosen before $A_{j,m}$ is observed and remain within these limits.

For each $m$, Clopper--Pearson gives $\Prob_0(q_m>q_m^+\mid\mathcal C_{\mathrm{edge}})\le\gamma/K$. No independence across events is needed. A union bound gives
\begin{equation}
\Prob_0(\mathcal A\mid\mathcal C_{\mathrm{edge}})\ge1-\gamma,
\qquad
\mathcal A=\{q_m\le q_m^+\text{ for }m=0,\ldots,K-1\}.
\label{eq:coverage-event}
\end{equation}
This is the meaning of simultaneous coverage: one event certifies all bounds used by the mixture.

\paragraph{Proposition.} Under the assumptions above, conditional on any calibration realization in $\mathcal A$, every $(M_{J,m})_{J\ge0}$ and their fixed mixture $(M_J)_{J\ge0}$ are nonnegative supermartingales with initial value one. Hence
\[
\Prob_0\!\left(\left.\sup_{J\ge1}M_J\ge\frac{1}{\alpha}\,\right|\,
\mathcal C=\mathsf c\right)
\le\alpha
\qquad\text{for every }\mathsf c\in\mathcal A.
\]

\paragraph{Proof.} Fix $m$. Since $A_{j,m}\in\{0,1\}$,
\[
E_{j,m}=
\begin{cases}
1-\lambda_{j,m}q_m^+ \ge 1-c>0, & A_{j,m}=0,\\
1+\lambda_{j,m}(1-q_m^+) \ge 1, & A_{j,m}=1.
\end{cases}
\]
Thus $M_{J,m}\ge0$. On $\mathcal A$, predictability and \eqref{eq:null} give
\begin{align*}
\E_0[E_{j,m}\mid\mathcal G_{j-1}]
&=1+\lambda_{j,m}\left(\E_0[A_{j,m}\mid\mathcal G_{j-1}]-q_m^+\right)\\
&\le1+\lambda_{j,m}(q_m-q_m^+)\le1.
\end{align*}
Since $M_{j-1,m}$ is nonnegative and measurable given $\mathcal G_{j-1}$,
\[
\E_0[M_{j,m}\mid\mathcal G_{j-1}]
=M_{j-1,m}\E_0[E_{j,m}\mid\mathcal G_{j-1}]
\le M_{j-1,m}.
\]
Each component is therefore a nonnegative supermartingale. Linearity of conditional expectation makes their fixed average $M_J$ a nonnegative supermartingale with $M_0=1$. Ville's inequality gives the conditional result. Finally,

\[
\Prob_0\!\left(\sup_J M_J\ge1/\alpha\right)
\le \Prob_0(\mathcal A^c)+
\Prob_0\!\left(\sup_J M_J\ge1/\alpha,\mathcal A\right)
\le\alpha+\gamma=\deltafa.
\]
\hfill$\square$

The proof does not assume anything about dependence among tokens within a response. It does assume that calibration responses provide binomial observations for each event and that their marginal prompt-response event bounds transfer to future rounds. If $q_m^+=\E_0[A_{j,m}\mid\mathcal G_{j-1}]$, then the corresponding factor has conditional mean one. A conservative upper bound changes equality to an inequality, which is why the implemented process is generally a supermartingale rather than a martingale. The uniform mixture costs at most $\log K$ in log evidence relative to its best component: a component reaching $K/\alpha$ is sufficient by itself to trigger the alarm. With $K=7$, this is a small fixed price for not selecting an event threshold on attack data.

\section{Experimental data and evaluation protocol}
\label{app:experimental-details}

\subsection{Threat model and seed access.}\label{subsec:threat-model} We assume a trusted verifier and a serving stack whose token-selection logic may be compromised. The attacker knows the model, prompt, logits, sampling rule, and detector, and may coordinate its token choices with a receiver to encode a payload. It cannot alter the verifier, its calibration data, or the recorded responses. Its only lever is the token returned at each step, so detection must come from the pattern of emitted tokens rather than (possibly) corrupted telemetry. 

We separate two seed-access regimes. A \emph{seed-blind} attacker chooses the next token to emit without seeing the verifier's sampling seed, replayed token, or seeded margin. This requires an architecture that keeps seed-derived randomness outside the compromised encoding component, such as an isolated sampler or a commit-then-reveal protocol. By contrast, a \emph{seed-aware} attacker owns the sampling seed and can reconstruct the verifier's sampling step before choosing a token, concentrating its payload where the replayed winner is nearly tied with the runner-up. The closer it matches the benign pattern of mismatches and replay scores, the less evidence the monitor can accumulate. We treat this stronger regime as a stress test rather than assuming that seed secrecy survives full server compromise.

\paragraph{Seed-aware attack.} At each position, the seed-aware attacker computes the seed-prescribed token $Y^*_{j,t}$ and the runner-up under the verifier's fixed-seed likelihood score, with score gap $d_t$. A position is eligible to encode bits when the runner-up has score at least $\tau=-5$ and $d_t\le d_{\max}$. Each eligible position offers a binary alphabet: $Y^*_{j,t}$ encodes $0$ and the runner-up encodes $1$; otherwise the attacker emits $Y^*_{j,t}$ without consuming a bit. Hence $d_{\max}$ controls how many positions offer one bit, not how many bits one eligible position carries. The sets are nested: positions eligible at $.02$ remain eligible at $.05$ and $.1$, which additionally admit wider gaps. If $K(d_{\max})$ of $N$ positions qualify, the attack carries $K(d_{\max})$ bits at rate $K(d_{\max})/N$ bits per token; increasing $d_{\max}$ therefore raises both the realized rate and exposure to detection. A four-candidate alphabet could instead map $00,01,10,11$ to four tokens, but that is a different attack: the implementation evaluated here always uses the top two. The decoder reconstructs this ordered pair from the emitted token and returned top-candidate logprobs. The binary code, payload distribution, and detector remain fixed across $d_{\max}$.

\paragraph{Seed-blind attack.} Unable to compute $Y^*_{j,t}$, the encoder forms a codebook from up to $16$ clean candidates within $0.5$ logits of the maximum, emitting the maximum if fewer than two qualify. Its size varies and is often not a power of two, so fixed bit labels waste choices. Arithmetic coding efficiently uses the full nonuniform codebook with model-proportional weights and is reconstructible from clean logprobs without the seed~\citep{meteor2021,imec2023}. It is therefore an efficiency choice: the seed-aware binary top-two code instead isolates $d_{\max}$. Seeded noise can reorder clean candidates, causing wide-margin mismatches. Rates are ideal channel-use estimates, not measured payload.

\paragraph{Decoding assumptions.} Both implementations assume that a colluding API client receives the emitted token and enough top-candidate logprobs to reconstruct each codebook and, for arithmetic coding, its weights. If the provider omits logprobs, or truncates them below the candidate-set size, this decoder cannot recover the payload. Some frontier interfaces do not expose the required metadata: Anthropic's Messages schema has no logprobs field, OpenAI's GPT-5.6 feature list does not advertise logprobs, and Google marks both logprob controls deprecated for Gemini 3.x~\citep{anthropic2026messages,openai2026gpt56,google2026gemini}. This blocks the implemented decoder on those interfaces, but not the sampler channel in principle: an attacker could use a lower-rate codebook recoverable from output tokens alone.

\subsection{Dataset.}\label{subsec:dataset} Prompts come from the training split of LMSYS-Chat-1M~\citep{zheng2024lmsys}. After applying each model's chat template, we retain unique conversations of at most $512$ prompt tokens. The main experiment uses $250$ even-indexed prompts for trusted calibration and $250$ odd-indexed prompts for testing. Each attack run and its benign control use the same odd prompts, so their only intended difference is whether payload bits are encoded. Each setting has one response per prompt, capped at $512$ output tokens. Table~\ref{tab:experimental-corpus} gives the scored token counts and Table~\ref{tab:experimental-config} the monitoring settings.

The four checkpoints cover three model families and two architectures: Llama-3.2-3B and Qwen3-32B are dense, while OLMoE-1B-7B and Qwen3-30B-A3B are mixture-of-experts models. This samples benign replay behavior across different families, nominal scales, and routing architectures, but does not isolate any one factor or constitute a comprehensive model survey.

\begin{table}[ht]
  \centering
  \caption{Scored output-token counts for the seed-aware attack. Attack counts span the three reported $d_{\max}$.}
  \label{tab:experimental-corpus}
  \footnotesize
  \setlength{\tabcolsep}{3.0pt}
  \begin{tabular}{@{}llrrr@{}}
    \toprule
    Model & Architecture & Calibration & Benign test & Attack range \\
    \midrule
    Llama-3.2-3B-Instruct       & dense & $44{,}201$  & $52{,}367$  & $51{,}839$--$52{,}547$ \\
    OLMoE-1B-7B-0125-Instruct  & MoE   & $78{,}985$  & $79{,}375$  & $78{,}837$--$80{,}248$ \\
    Qwen3-32B                  & dense & $113{,}613$ & $111{,}679$ & $112{,}138$--$112{,}570$ \\
    Qwen3-30B-A3B-Instruct     & MoE   & $71{,}412$  & $69{,}855$  & $69{,}616$--$70{,}473$ \\
    \bottomrule
  \end{tabular}
\end{table}

\begin{table}[ht]
  \centering
  \caption{Experimental and monitoring settings.}
  \label{tab:experimental-config}
  \footnotesize
  \setlength{\tabcolsep}{4pt}
  \begin{tabular}{@{}ll@{}}
    \toprule
    Setting & Value \\
    \midrule
    Seed-aware attack & seeded winner and runner-up; one bit per eligible token \\
    Maximum score gap & $d_{\max}\in\{.02,.05,.1\}$ \\
    Eligibility & runner-up predicted safe and score gap $d_t\le d_{\max}$ \\
    Margin construction & six fixed percentile levels: $2,5,10,20,40,70\%$ \\
    Margin events & $K=7$ nested margin thresholds \\
    Benign split per model & $125$ edge fit; $125$ event calibration; $250$ matched test \\
    e-process & $c=0.5$; $\alpha=0.005$; $\gamma=0.005$; threshold $200$ \\
    Rinberg et al. baseline & $\sigma=1$; $\tau=-5$; rank cutoff $R=8$ \\
    Escalation policy & first response containing a dangerous token \\
    Monitoring protocol & $1{,}000$ resampled streams of $250$ responses per setting \\
    \bottomrule
  \end{tabular}
\end{table}

\begin{table}[ht]
  \centering
  \caption{Fitted margin thresholds ($125$ edge-fitting responses). $b_0=0$, whereas the last cell includes margins above $b_6$.}
  \label{tab:fitted-cutoffs}
  \footnotesize
  \setlength{\tabcolsep}{3.8pt}
  \begin{tabular}{@{}lrrrrrr@{}}
    \toprule
    Model & $b_1$ & $b_2$ & $b_3$ & $b_4$ & $b_5$ & $b_6$ \\
    \midrule
    Llama-3.2-3B  & .0425 & .1180 & .2374 & .4713 & 1.0474 & 2.2907 \\
    OLMoE-1B-7B   & .0387 & .1000 & .2072 & .4349 & .9791 & 2.1589 \\
    Qwen3-32B     & .0454 & .1098 & .2252 & .4648 & 1.0206 & 2.2092 \\
    Qwen3-30B-A3B & .0464 & .1165 & .2354 & .4884 & 1.0753 & 2.2991 \\
    \bottomrule
  \end{tabular}
\end{table}

\paragraph{Data generation and verification.} Generation and replay use the same bfloat16 checkpoint with temperature $1$, top-$k=50$, top-$p=0.95$, sampling seed $42$, and concurrency $1$. Replay is run separately and records the mismatch indicator, seeded margin, fixed-seed likelihood score, and raw-logit rank. The payload is a pseudorandom bit stream with seed $1234$, separate from the sampling seed.

\paragraph{Calibration and evaluation.} For each model, the $250$ even-prompt benign responses are divided by a fixed whole-prompt split into $125$ responses for margin-threshold fitting and $125$ for event bounds. For each of the $K=7$ events, we use the Clopper--Pearson bound at error $\gamma/K$. We set $\alpha=\gamma=0.005$ and $c=0.5$, giving alarm threshold $200$ and the $0.01$ bound under the prompt-level null and calibration-transfer assumption. A separate benign run on the $250$ odd prompts is the matched benign control set. Each attack setting uses those same odd prompts, and we form $1{,}000$ streams by resampling complete responses with replacement. For Table~\ref{tab:compromise} and Figures~\ref{fig:capacity-detection} and~\ref{fig:capacity-latency} both detectors run at their prespecified alarm rule (e-process threshold $200$, Rinberg et al. alarms at the first dangerous token), and we record each stream's first-alarm response and whether it alarms by response $250$. Evidence is updated once after each complete response. These bootstrap streams quantify sensitivity to resampling and response order within one fixed $250$-response test pool, they are not independent model runs or confidence intervals for a new deployment. Table~\ref{tab:fitted-cutoffs} reports the fitted margin values.

\subsection{Baseline \texorpdfstring{\citep{rinberg2025}}{(Rinberg et al.)}}\label{subsec:rinberg} The baseline scores each served token by how consistent it is with the committed seed. Holding that seed, the verifier recomputes the exact Gumbel perturbation the sampler added to the logits at that step and finds the token the seeded sampler would have selected, $Y^*_{j,t}$. Writing $\Delta^{\mathrm{srv}}_{j,t}\ge0$ for the resulting gap on the perturbed logits between $Y^*_{j,t}$ and the served token, the \emph{fixed-seed likelihood score} is
\begin{equation}
\GLS_{j,t}=\log\Phi\!\left(-\Delta^{\mathrm{srv}}_{j,t}/(\sigma\sqrt2)\right),
\label{eq:gls}
\end{equation}
where $\Phi$ is the standard normal cumulative distribution function and $\sigma$ scales the assumed noise. When the served token is the prescribed winner the gap is zero and the score takes its maximum $\log\tfrac12\approx-0.69$. A token far down the seeded order has a large gap and a strongly negative score. Intuitively, the score is high when the served token is what the seed would have produced and falls as the two disagree by more.

On top of this score, \citet{rinberg2025} propose a fixed three-tier rule: a token is \emph{safe} when its score is at least $\tau$, a lower-scoring token is \emph{suspicious} when its zero-indexed raw-logit rank is at most $R$ and \emph{dangerous} when its rank is greater than $R$. We use the verifier's $\sigma=1$ scale and the recommended threshold from the paper at $\tau=-5$, together with the rank cutoff $R=8$ highlighted in their results. The authors explicitly leave pooling to the response level and escalation to the deployment. We therefore define the hard-alarm policy as the first response containing at least one dangerous token. This policy has no claimed stream-level error guarantee.

Section~\ref{sec:experiments} explains why its rank gate alarms on benign traffic yet misses the seed-blind attack. The e-process instead tests whether mismatches occur at wider seeded margins more often than under benign serving.

\section{Detailed experimental results}
\label{app:detailed-results}

This appendix supports the three main findings of Section~\ref{sec:experiments} with the full monitoring detail: (1) every resampled seed-blind stream alarms quickly; (2) seed-aware detection depends on rate and the model's benign margin profile; and (3) no held-out benign e-process stream crosses, whereas the hard-alarm baseline crosses frequently. All results below use the data splits of Appendix~\ref{app:experimental-details} and one shared protocol: for each model and setting, $1{,}000$ audit streams of $250$ complete responses, sampled with replacement from the test prompts, with each detector running its prespecified alarm rule.

\begin{figure}[H]
  \centering
  \includegraphics[width=\linewidth]{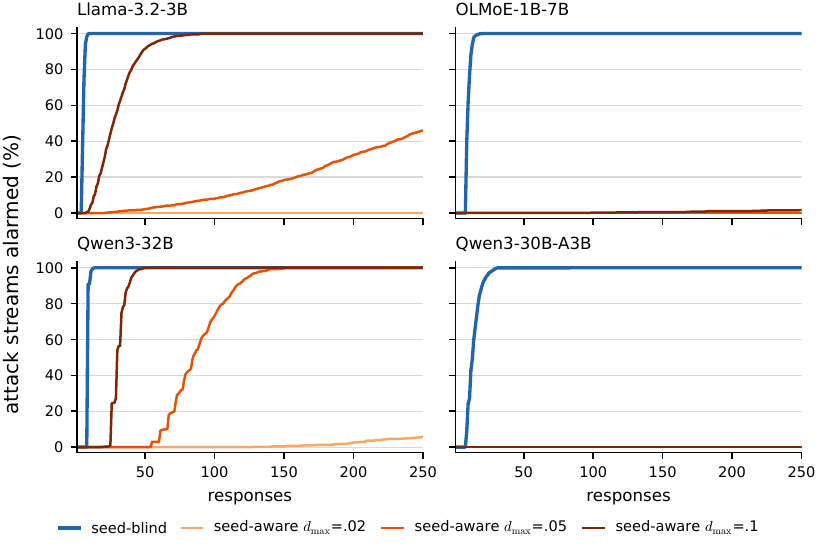}
  \caption{Fraction of attack streams alarmed by each response ($1{,}000$ bootstrap streams of $250$ responses), seed-blind versus seed-aware at $d_{\max}\in\{.02,.05,.1\}$.}
  \label{fig:attack-ecdf}
\end{figure}

Figure~\ref{fig:attack-ecdf} plots the fraction of attack streams that have alarmed by each response. Against the seed-blind attack, every resampled stream on every model alarms, typically within the first fifteen responses (medians $6$--$14$), and no matched benign stream alarms. Resampling response order $1{,}000$ times shows that this result is not tied to the recorded order. The seed-aware attack behaves differently: on the dense models the alarms arrive earlier as $d_{\max}$ grows, while on the two mixture-of-experts models the curves stay essentially at zero (at most $1.7\%$ of streams by response $250$).

\begin{figure}[ht!]
  \centering
  \includegraphics[width=\linewidth]{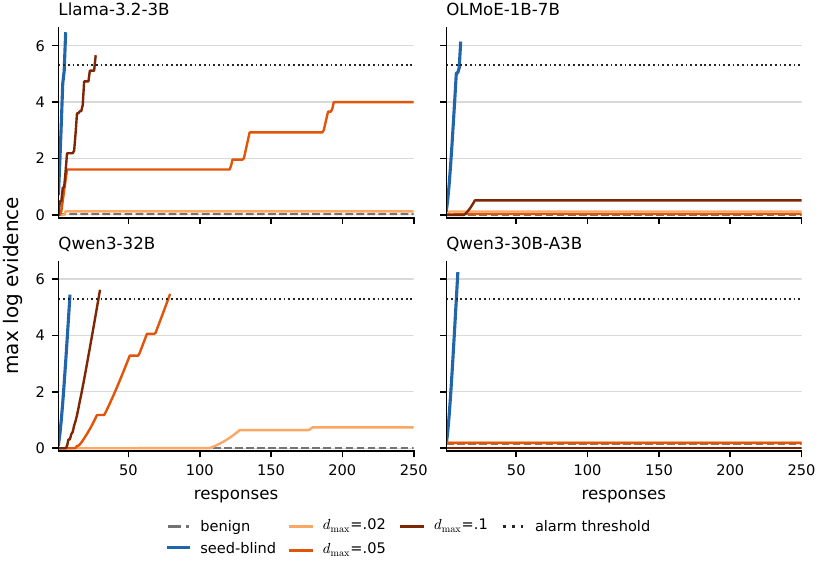}
  \caption{Accumulated log evidence for the benign control, the seed-blind attack, and the seed-aware attack at each $d_{\max}$. Curves stop at the alarm threshold $\log 200$.}
  \label{fig:two-choice-wealth}
\end{figure}

First, we examine the evidence paths of each model and attack setting. Figure~\ref{fig:two-choice-wealth} shows one recorded-order path per setting. The seed-blind paths cross within the first few responses. Seed-aware paths cross only at detectable settings, whereas the matched benign paths remain below the alarm threshold. Table~\ref{tab:compromise} and Figure~\ref{fig:attack-ecdf} summarize the bootstrap streams.

Detectability is model dependent. Figure~\ref{fig:capacity-detection} follows from the benign floor of Table~\ref{tab:benign-flips} and Figure~\ref{fig:margin-law}. The two mixture-of-experts models produce several times more benign mismatches, including at wide seeded margins, so their calibrated benign event bounds are high. A seed-aware attacker that limits payload swaps to near-ties can then sit beneath that floor, leaving little excess for the monitor to accumulate. Whether this attack is detectable is therefore as much a property of the model--serving stack's benign behavior as of the monitor. Appendix~\ref{app:margin} explains why the four-model comparison does not establish an architecture effect.

\begin{figure}[H]
  \centering
  \includegraphics[width=\linewidth]{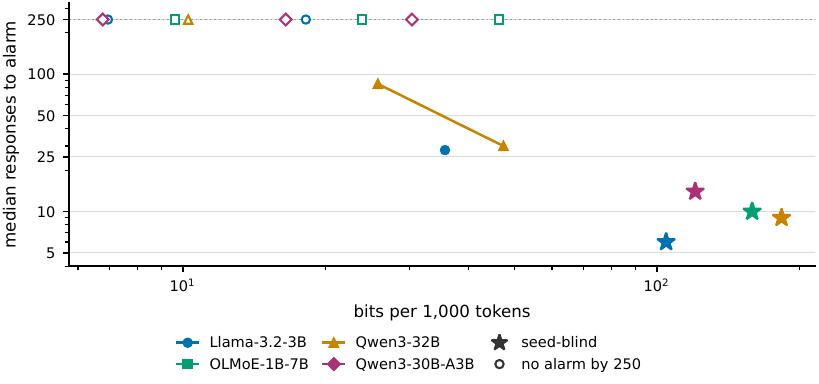}
  \caption{Median response at first e-process alarm by realized rate ($1{,}000$ streams. Medians over all streams, censored at response $250$). Stars: seed-blind attack; open markers: no alarm by response $250$.}
  \label{fig:capacity-latency}
\end{figure}

Figure~\ref{fig:capacity-latency} condenses the same $d_{\max}$ sweep into one panel: the median response at first e-process alarm against the realized exfiltration rate, counting streams that never alarm as censored at the horizon, so settings whose median stream never alarms sit as open markers on the top line. Across the seed-aware settings that are detected, alarms arrive sooner at higher realized rates. The evaluated seed-blind codec operates at a higher rate and alarms within the first dozen responses. This is an empirical trade-off for these codecs, not proof that throttling is an attacker's only evasion strategy.

\begin{figure}[H]
  \centering
  \includegraphics[width=\linewidth]{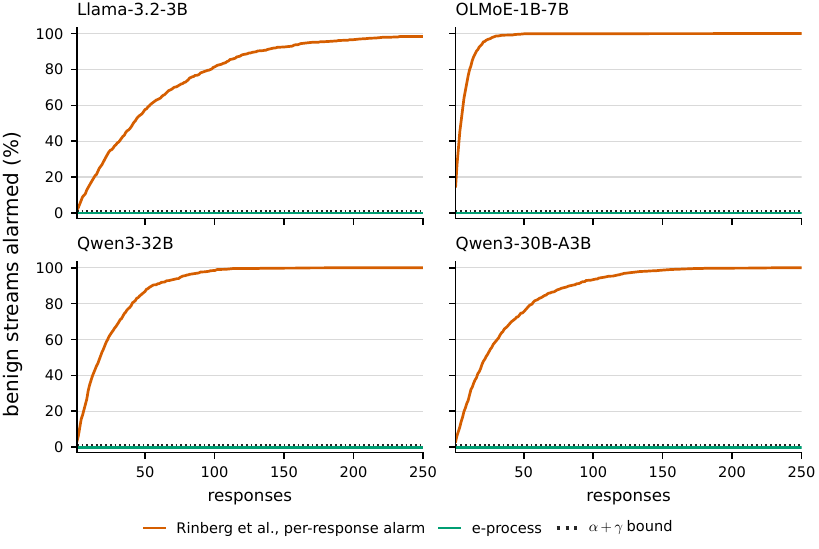}
  \caption{Cumulative false-alarm rate on benign streams by response ($1{,}000$ bootstrap streams). Dotted: the $\alpha+\gamma=1\%$ bound.}
  \label{fig:repeated-fpr}
\end{figure}

At last, looking in depth at the benign control set, we can observe how on the held-out benign data the largest recorded-order log evidence over $250$ responses is $0.15$ (Qwen3-30B-A3B), well below the alarm line $\log 200\approx5.3$, and none of the $1{,}000$ resampled benign streams crosses on any model. Figure~\ref{fig:repeated-fpr} shows that, when checked response by response on the same streams, the hard-alarm baseline's cumulative benign crossing rate approaches $100\%$, whereas the e-process has no crossings. Theorem-level anytime control comes from Section~\ref{sec:method} and Appendix~\ref{app:validity}.

\end{document}